\documentclass[%
 reprint,
superscriptaddress,
 amsmath,amssymb,
 aps,
pra,
]{revtex4-2}

\usepackage{graphicx}
\graphicspath{{./figs/}}
\usepackage{caption}

\usepackage{subcaption}
\usepackage{dcolumn}

\usepackage{bm}
\usepackage{algorithm,algcompatible}
\usepackage[usenames,dvipsnames]{color} 

\usepackage{ifthen} 
\newcommand{\INITIAL}{\item[\algorithmicinitial]}
\newcommand{\algorithmicinitial}{\textbf{Initial:}}
\newcommand{\RETURN}{\item[\textbf{Return:}]}

\def\shownoteal{1} 
\newcommand{\nal}[1]{\ifthenelse{\shownoteal=1}{\textcolor{red}{[[#1]]}}{}}

\begin{document}

\preprint{APS/123-QED}

\title{Analog quantum variational embedding classifier}

\author{Rui Yang}
%\email{r249yang@uwaterloo.ca}
\address{
 Institute for Quantum Computing, and Department of Physics and Astronomy, University of Waterloo, Waterloo,
Ontario, N2L 3G1, Canada\\
}

 \author{Samuel Bosch}
 
 \address{
 Massachusetts Institute of Technology, 77 Massachusetts Avenue, Cambridge, MA 02139, USA\\
}

\author{Bobak Kiani}
\address{
 Massachusetts Institute of Technology, 77 Massachusetts Avenue, Cambridge, MA 02139, USA\\
}

\author{Seth Lloyd}
\address{
 Massachusetts Institute of Technology, 77 Massachusetts Avenue, Cambridge, MA 02139, USA\\
}

\author{Adrian Lupascu}
%\email{alupascu@uwaterloo.ca}
\address{
 Institute for Quantum Computing, and Department of Physics and Astronomy, University of Waterloo, Waterloo,
Ontario, N2L 3G1, Canada\\
}

\date{\today}

\begin{abstract}

 Quantum machine learning has the potential to provide powerful algorithms for artificial intelligence. The pursuit of quantum advantage in quantum machine learning is an active area of research. For current noisy, intermediate-scale quantum (NISQ) computers, various quantum-classical hybrid algorithms have been proposed. One such previously proposed hybrid algorithm is a gate-based variational embedding classifier, which is composed of a classical neural network and a parameterized gate-based quantum circuit. We propose a quantum variational embedding classifier based on an analog quantum computer, where control signals vary continuously in time: our particular focus is an implementation using quantum annealers. In our algorithm, the classical data is transformed into the parameters of the time-varying Hamiltonian of the analog quantum computer by a linear transformation. The nonlinearity needed for a nonlinear classification problem is purely provided by the analog quantum computer through the nonlinear dependence of the final quantum state on the control parameters of the Hamiltonian. We performed numerical simulations that demonstrate the effectiveness of our algorithm for performing binary and multi-class classification on linearly inseparable datasets such as concentric circles and MNIST digits. Our classifier can reach accuracy comparable with the best classical classifiers. We find that the performance of our classifier can be increased by increasing the number of qubits until the performance saturates and fluctuates. Moreover, the number of optimization parameters of our classifier scales linearly with the number of qubits. The increase of number of training parameters when the size of our model increases is therefore not as fast as that of neural network. Our algorithm presents the possibility of using current quantum annealers for solving practical machine-learning problems, and it could also be useful to explore quantum advantage in quantum machine learning.

\end{abstract}

\maketitle

\section{\label{sec:level1}Introduction:\protect
}

Recent progress in the field of quantum computation has led to attaining several important milestones. Quantum supremacy has been demonstrated in a few platforms~\cite{Arute_2019, USTC2021, Xanadu_2022}. Quantum processors with tens of qubits have been implemented and become available for operation via clouds \cite{ibmq}. Quantum error correction using the surface code~\cite{Fowler_surface_code} has been demonstrated in several experiments~\cite{uvzhao_qecc, Wallraff_qecc, googleQECC2021}.

Quantum computers have the potential to solve certain problems significantly more efficiently than classical computers. The original motivation for using quantum computers is to efficiently simulate quantum mechanical systems such as molecules~\cite{feynman1981simulating}. Besides this quantum advantage foreseeable in quantum chemistry, a quantum advantage could also be found in other computation problems. For example, Shor's quantum algorithm can factorize large numbers efficiently~\cite{shor1999polynomial}, thereby posing a threat to the widely-used RSA encryption system upon which the modern financial system is built. Grover's algorithm can speed up unstructured database search~\cite{grover1997quantum}.

A promising area of application for quantum computers is machine learning. The field of machine learning has developed at a fast pace in recent years. Since 2011, the progress made in neural-network algorithms and large-scale classical computing hardware, such as graphical processing units (GPU), has led to increased performance of machine learning for many applications. A notable example is image classification, where such algorithms perform better than humans. Machine learning algorithms have a wide range of profitable industry applications, such as recommendation system~\cite{2015NatureDL, ComputerVision1}. Whether quantum computers can contribute to the machine learning community by providing quantum advantage is an intriguing question~\cite{2017NaturQML}.

To explore the potential advantage brought about by quantum machine learning, many quantum machine learning algorithms have been proposed~\cite{2017NaturQML, Seth_2014NP, lloyd2016quantum, PRLQSVM, qgan_seth}. Some of the quantum machine learning algorithms have been shown to have advantages over their classical counterparts~\cite{Seth_2014NP, lloyd2016quantum, PRLQSVM}. Many proposed algorithms fall into two categories: classification~\cite{2019NatureIBM, PRLQSVM} or generative tasks~\cite{qgan_seth}. For classification tasks, several algorithms based on variational embedding~\cite{2019NatureIBM}, quantum kernels~\cite{2019NatureIBM}, or quantum support vector machines~\cite{PRLQSVM} have been proposed. The mechanism behind many quantum classification algorithms is based on finding a simplified separation boundary between different classes after nonlinearly mapping the original data into a new space that usually has a much higher dimension ~\cite{SethLloyd, Murphy_clustering}.

We now discuss recent progress on variational embedding classifiers. Havl{\'\i}{\v{c}}ek {\it et al.} proposed and demonstrated a variational embedding algorithm for binary classification tasks~\cite{2019NatureIBM}, which embeds classical data into quantum states of a high dimensional Hilbert space through parameters of quantum gates, and uses quantum measurements to classify the embedded quantum states. This approach focuses on finding an operator whose expectation value with respect to the states corresponding to embedded data is different for the two classes. Specifically, the two classes lead to positive and negative expectation values, respectively. This decision operator effectively defines a hypersurface in the multi-qubit Hilbert space used to separate embedded quantum states corresponding to two classes. Lloyd {\it et al.} proposed a more general version of the variational method~\cite{SethLloyd}, which uses a general decision operator defined with density matrices from collections of embedded data. This algorithm has a training procedure that optimizes the embedding to maximize the distance between embedded quantum states with different labels. In this approach, a neural network is used to transform the classical data into a dataset with different dimensionality. A parameterized quantum circuit is introduced where the rotation angles of the gates are the circuit parameters. Each data can be mapped into a final evolved quantum state of a multi-qubit Hilbert space by filling some of the rotation angles of the quantum circuit with the transformed data values. The rest of the quantum circuit parameters are knobs used to maximize the distance between the density matrices formed by states of each class. After maximizing the distances, the classification can be done using a simple decision operator expectation value metric to find the label. One open question regarding this algorithm is with regards to the role of the quantum part in the operation of the algorithm since the nonlinearity in the neural network is sufficient by itself for performing classification.

To elucidate the source of nonlinearities in a hybrid classifier and establish whether the quantum part alone can do the ``heavy-lifting'' for realizing a nonlinear classifier, we propose and investigate a new quantum variational embedding classifier based on an analog quantum computer, whose control fields vary continuously in time. We focus on an implementation corresponding to the transverse field Ising model used in quantum annealing; however, this approach can be explored with other types of Hamiltonians. In our algorithm, the neural network is replaced with a simple linear transformation, and the variational quantum circuit composed of gates is replaced with an analog quantum computer with direct control of the continuously varying Hamiltonian parameters. The nonlinear mapping of the classical data to a high dimensional density matrix is now realized with the analog quantum computer. The mapping can be regarded as a point in a $2^n \times 2^n$ dimensional space defined on a complex number domain, where $n$ is the number of qubits. Efficient classification is achieved when the data corresponding to distinct labels form separate clusters in Hilbert space. Separability is analyzed in terms of the distinguishability of density matrices averaged over the states corresponding to different labels. To form separate clusters, the $L_2$ (Hilbert-Schmidt) distance~\cite{SethLloyd} between the averaged density matrices is maximized in the training stage by adjusting the parameters in the linear transformation that converts classical data into the parameters of the analog quantum computer. In the classification stage, a simple classifier based on distance metric is used, yielding the predicted label for a new dataset as the label of the closest averaged density matrix obtained at the training stage. This distance-based strategy is also known as nearest centroid classification~\cite{nearest_centroid}. This distance metric could be obtained purely with a quantum circuit~\cite{SethLloyd}. We find that our algorithm can classify linearly inseparable datasets~\cite{NooriBarry} with high accuracy. The dependence of performance on the number of qubits shows that increasing the number of qubits can boost performance. Our algorithm opens up the possibility to use quantum annealers~\cite{Hauke_2019} for solving practical machine-learning problems.

\section{Algorithm}

In this section, we present our algorithm, which realizes a quantum variational embedding classifier on multi-class datasets. An important distinguishing feature of our algorithm, when compared to previous work, is that it employs control of the quantum system that is done via control of its Hamiltonian, in contrast with previous work where control is done based on quantum gates, as developed in the context of gate-based quantum computation model. 

In this work, we focus on the implementation of an analog variational embedding using a quantum annealer. Quantum annealing is an example of an analog quantum computation~\cite{NooriBarry}. In quantum annealing, the initial system Hamiltonian is a transverse field Hamiltonian, and the initial state is the ground state of this Hamiltonian. The Hamiltonian is continuously deformed, reaching at the end of the evolution an Ising form. This approach has been developed and explored in connection with prospects for solving hard computational problems. 

While we retain the key elements of quantum annealing, such as initial ground state preparation and continuous transverse to Ising Hamiltonians interpolation, our work focuses on optimizing evolution for classification.   Perhaps most important, we do not assume that the quantum annealer remains in its ground state throughout the analog computation: that is, the quantum annealer performs {\it diabatic} quantum annealing, in which the continuous time control fields drive it to a non-equilibrium final state.   We emphasize that our algorithm can be implemented with other types of Hamiltonians, such as Hamiltonians used in gate-based quantum computers: our method applies to any quantum information processor that is controlled by continuously time-varying fields.

The Hamiltonian of the quantum annealer used in our algorithm takes the form,

\begin{align} \label{eq:1}
    H(t)  =  (1-s) H_{0}(s) + s H_{1}(s) \nonumber\\+ H_{add}(s) ,
\end{align}
where $s=t/t_{max}$, with $t$ the time and $t_{max}$ is the total evolution time.

The initial $H_{0}$ and final $H_{1}$ Hamiltonians are given by  
\begin{align} \label{eq:2}
H_{0} =      \sum_{i} h_{x_i}\sigma_{x_i} 
\end{align}
and
\begin{align} \label{eq:3}
    H_{1} =      \sum_{i} h_{z_i}\sigma_{z_i} + \sum_{i,j}J_{ij}   \sigma_{z_i}  \sigma_{z_j} 
\end{align}
where $\sigma_{x_i}$ and $\sigma_{z_i}$ are Pauli matrices.

The additional Hamiltonian $H_{add}$  vanishes at the beginning and the end of the evolution ($H_{add}(0) = H_{add}(1) = 0$) and is given by: 
\begin{align} \label{eq:4}
    H_{add}(s) =  \sum_{i} P_{z_i}(s)  \sigma_{z_i}  +  \sum_{i} R_{x_i}(s) \sigma_{x_i} \nonumber\\
    + \sum_{i,j} V_{ij}(s)  \sigma_{z_i}  \sigma_{z_j} 
\end{align}

The time-dependent coefficients in front of Pauli matrices (schedules) take the following form:
\begin{align}
    P_{z_i}(s) =  \sum_{k} c_{z_i,k} \sin( (k+1) \pi s),  \\ 
    R_{x_i}(s) =  \sum_{k} c_{x_i,k} \sin((k+1) \pi s), \text{and}  \\    
    V_{ij}(s) =  \sum_{k} c_{ij,k} \sin((k+1) \pi s) .     
\end{align}
These forms are complete Fourier series expansions consistent with the cancellation condition at $s=0$ and $s=1$. The Fourier expansion waveform ansatz of the control fields used for quantum optimal control is an example of chopped random basis ({\it CRAB}) \cite{CRAB1, CRAB2}. Previous studies on CRAB show the optimization landscape of CRAB ansatz is good for trainability~\cite{CRAB1, CRAB2}, and as few as three Fourier terms in the Fourier ansatz are enough for a good performance \cite{CRAB1, CRAB2}.

We emphasize that our Hamiltonian is assumed to have full independent control of the initial transverse field and final Ising Hamiltonians, as well as of the time dependence of the additional Hamiltonian. This model is consistent with a recently developed platform for coherent quantum annealing~\cite{sergey_annealer}.   

\begin{figure*}[t]
	\begin{center}
		\includegraphics[width=2\columnwidth]{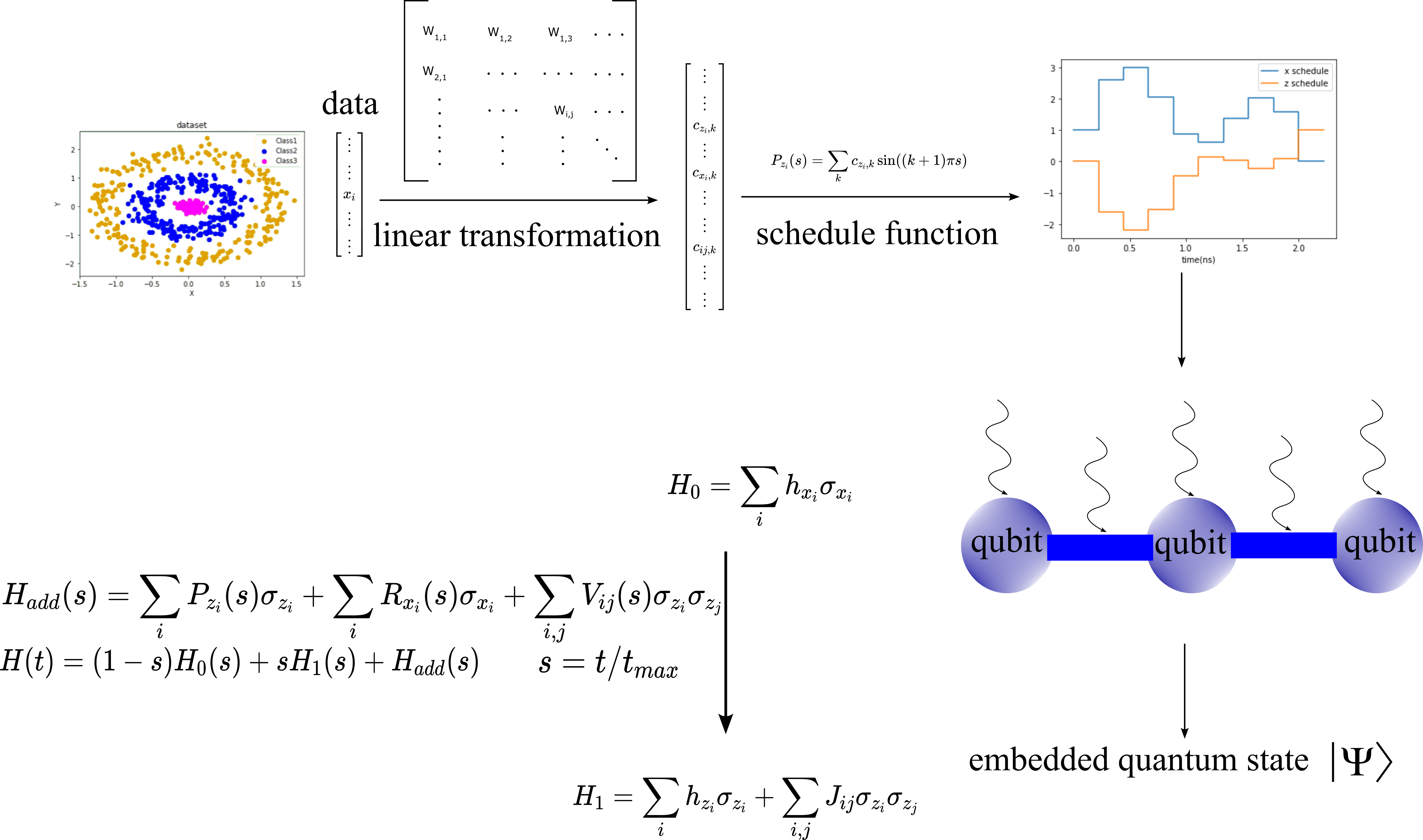} 
    \end{center}
    \caption{Illustration of our analog quantum variational embedding classifier. The balls placed in a line represent a multi-qubit quantum annealer. The connection lines represent couplings. The wavy arrows represent time-dependent driving on the qubits and couplings. The Hamiltonian of the annealer starts with $H_{0}$ and gradually changes to $H_{1}$, driven by a time-dependent $H_{add}$ (see equations (1) to (7)). The time-dependent coefficients (the schedule) of the Pauli terms of $H_{add}$ control the evolution, see equations (4) to (7). The schedule is expressed with a summation of sin terms to meet a vanishing boundary condition at the beginning and end of the evolution, see equations (5) to (7). The data is transformed into the parameters $c_{z_i,k}$, $c_{x_i,k}$ and $c_{ij,k}$ by a linear transformation matrix to define the schedule of the quantum annealer, see equations (5) to (9). }\label{fig:QClassifier}
\end{figure*}

The connection between the data in the classification problem and the Hamiltonian is made as a linear transformation from the data to the system parameters, such as the Fourier coefficients of the additional Hamiltonian, as illustrated in  Fig.~\ref{fig:QClassifier}. The configuration we use for this Hamiltonian is a simple one-dimensional configuration with nearest-neighbor ZZ coupling. Among all the possible configurations, this is the simplest configuration for building quantum computers and therefore is a good start for testing quantum algorithms.

In our algorithm, classical data is represented by a vector of dimension $d$, which is mapped into the parameters of an annealing Hamiltonian. The Hamiltonian is used to evolve a multi-qubit ($n$-qubit) quantum system; see Fig.~\ref{fig:QClassifier} and Algorithm~\ref{alg:architecture} for the illustration. For a $d$-dimensional data $\hat{X} = [x_{1},..., x_{d}]$ which represents a point in a $d$-dimensional space, it can be transformed to the schedule parameters with the following linear transformation:

\begin{align} \label{eq:8}
\bm{\hat{V}_{c}}  =
\bm{\hat{W}}
\bm{\hat{X}^{T}}
\end{align}

where
\begin{align} \label{eq:8b}
\bm{\hat{V}_{c}} = 
\begin{bmatrix}
\vdots  \\
c_{zi_k} \\
\vdots \\
c_{xi_k} \\
\vdots \\
\vdots \\
c_{ij_{k} } \\
\vdots\\
\end{bmatrix}
\end{align}
and $\bm{\hat{W}}$ is a matrix with a size of $(n_{s}\times m, d)$. $n_{s}$ is the number of sins in the ansatz, $m=2n + (n-1) $ is the number of Pauli terms in the Hamiltonian ($n$ is the number of qubits in a chain), $d$ is the dimension of the data. $n_{s}\times m \times d$ is hence the number of optimization parameters in this case.

For each classical data point, after obtaining the relevant Hamiltonian parameters through a linear transformation, a corresponding state is calculated based on the evolution under the Hamiltonian. This process is repeated for a set of classical data points drawn randomly from the data. Density matrices are formed by combining the states corresponding to each point within a certain class. The $L_2$ distance is maximized in the training stage by adjusting the linear transformation $\bm{\hat{W}}$, which converts the raw data into schedule parameters, as outlined in Algorithm~\ref{alg:architecture}. The absolute loss for the optimization or training is therefore based on $L_2$ distance. The entries ($\hat{W}_{ij}$) of the linear transformation are therefore taken to be the parameters to be adjusted to maximize the distances/minimize the loss. In the classification stage, the predicted label for a new dataset is the label of the closest average density matrix.

The parameters $h_{z_i}$, $h_{x_i}$ and $J_{ij}$ in equations (\ref{eq:2}) and (\ref{eq:3}) determine the initial and final energy levels of the Hamiltonian. They can be treated with the same footing as the Fourier coefficients -- $h_{x_i}$, $h_{z_i}$ and $J_{ij}$ parameters of $H_{0}$ and $H_{1}$ can be transformed from the data by a linear transformation matrix, i.e., $h_{x_i}$, $h_{z_i}$ and $J_{ij}$ can be appended to $\hat{V}_{c}$. These parameters could also be fixed, e.g., taking on values to make the initial and final Hamiltonian of the annealer non-degenerate. The $t_{max}$ is a fixed parameter, representing the total evolution time.

For multi-class classification, in order to simultaneously maximize the pair-wise distances, we define the absolute loss as the product of the pair-wise distance between an arbitrary pair of density matrices from collections of embedded data (see Algorithm~\ref{alg:architecture}). The definition of loss is not unique to multi-class classification. Essentially, any meaningful loss is acceptable, e.g., we can also define the loss absolute as a summation of the pair-wise distance between any pair of density matrices from collections of embedded data (which gives an algorithm with similar performance). Compared with previous papers on variational embedding classifiers, we extend the definition of loss to handle multi-class classification situations.

\begin{figure}[t]
	\begin{center}
		\includegraphics[width=1\columnwidth]{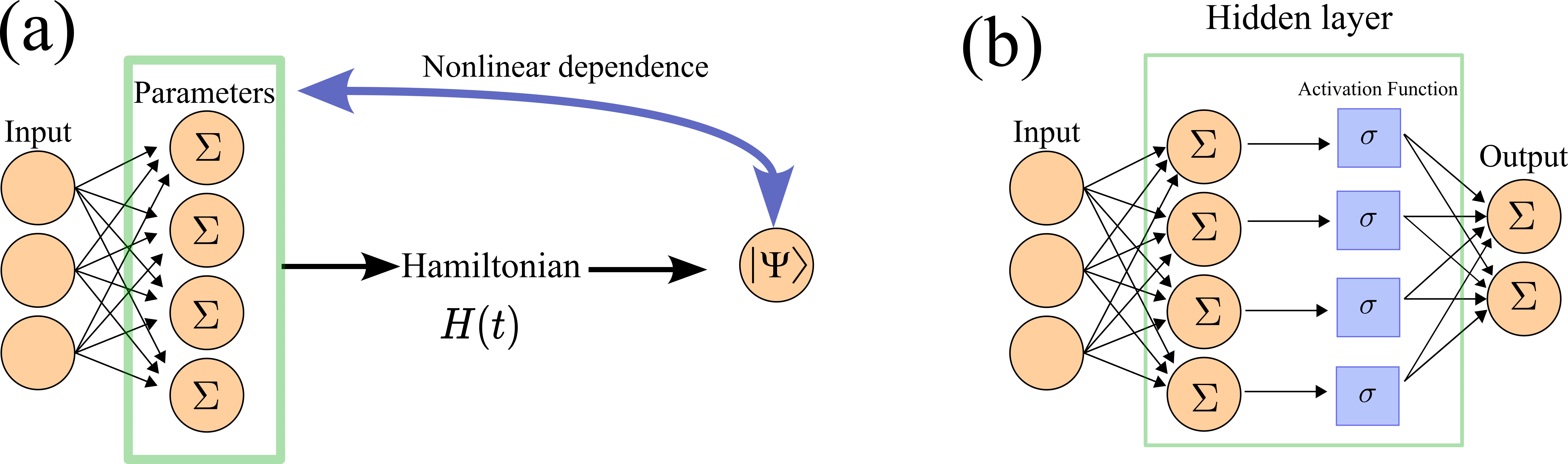} 
    \end{center}
    \caption{Comparing nonlinearity in our classifier and a neural network. (a)The nonlinearity in our classifier is from the nonlinear dependence between the quantum state and the parameters of the quantum system; (b)The nonlinearity in a neural network is from the activation function (such as a sigmoid function)}\label{fig:ComparingNonlinearity}
\end{figure}

Given that neural networks are a common way to implement classification, we discuss the comparison of our algorithm with a neural network. This comparison is illustrated in Fig.~\ref{fig:ComparingNonlinearity}, which shows a comparison of the nonlinearity in quantum variational circuits and a neural network. In a classical neural-network classifier, there are two stages: the first stage is a linear transformation which maps a point in the original data space into another point in another space. The next stage is a nonlinear transformation, which consists of a nonlinear distortion to the data points. It is this nonlinear distortion making the distorted mapped data points separate, and a simple boundary can be drawn between different classes. 
In our algorithm, the first stage is still linearly mapping the raw data into points in another space. The second stage is a nonlinear mapping into a $2^n \times 2^n$-dimensional space, taking advantage of the nonlinear dependence of the final state of a quantum system on the schedule parameters. In a quantum system, the mapping between the initial and final quantum state is linear since the evolution is from a unitary evolution. However, the mapping between the schedule parameters and the final state is nonlinear. This is the source of nonlinearity in our algorithm. The approach we use is similar to that of a classical neural network in that the raw data in the original data space is mapped into another space (here is the Hilbert space for our classifier) after being linearly transformed and nonlinearly distorted. The classification is performed on transformed data in the new space.

In the characterization of our algorithm, we use numerical simulations to calculate the evolved quantum states. In the numerical simulation, discrete time steps are used (see Fig.~\ref{fig:QClassifier}). In each time step, a constant Hamiltonian evolves the system, and the unitary evolution operator is calculated by direct matrix exponentiation: $U(t_n) = \exp{(\frac{-i}{\hbar}H(t_{n}) \delta t) } $ ($H(t_{n})$ is the Hamiltonian at time $t_{n}$, $\delta t$ is the time step). The total evolution is from the product of these unitary evolution operators: $U(t_{max}) = \prod\limits_{n} \exp{(\frac{-i}{\hbar}H(t_{n}) \delta t) } $. In the simulated classifier, to maximize the distance (with the linear transformation parameters as the control knobs), we use Pytorch's autograd~\cite{pytorch} feature to perform an efficient gradient descent optimization. Autograd~\cite{pytorch} keeps a record of tensors and all executed operations, and the resulting new tensors in a computational graph whose leaves are the input tensors and roots are the output tensors. By tracing this graph from roots to leaves, we can automatically compute the gradients using the chain rule. The whole simulation in this study is written with Pytorch objects with the autograd features, which is essential to propagate the gradient backward toward tens or even hundreds of control knob parameters.

\begin{algorithm}[H]
  \caption{Analog quantum variational embedding classifier algorithm}
  \label{alg:architecture}
  \begin{algorithmic}[1]
    \INITIAL
      Initialize parameters in the linear transformation matrix $\bm{\hat{W}}$;
      Initialize system state;

    \FOR{ iterations $j=1,Epochs$}

        \FOR{label iteration $i = 1,L_{classes}$}
        
            \FOR{sample in class $i$}
            
                    \STATE Convert the sample to parameters of the system Hamiltonian by equation (8), then, define the system Hamiltonian (equation (1)); Evolving the initial state of the system to get a density matrix for the sample
                    \STATE Add the density matrices together

            \ENDFOR
            
            \STATE Averaging: taking the summation of the density matrix divided by the number of samples in class $i$
            as an average density matrix $M_{i}$ for label i

        \ENDFOR
    
    \STATE Calculate the $L_2$ distance $D_{ij} = Tr( (M_{i} - M_{j})^2 )$ between the average density matrices $M_{i}$ and $M_{j}$

    \STATE Perform a gradient descent on the loss function $Loss= -\prod_{ij} D_{ij}$ (for a binary classification task, $Loss= -D_{ij}$; $-\prod_{ij} D_{ij}$ is an extended loss for a multi-class classification task, other loss definition will also work) with respect to the parameters in the linear transformation $\bm{\hat{W}}$.
    \STATE Update the parameters in the linear transformation $\bm{\hat{W}}$

    \ENDFOR
    \RETURN Parameters in the linear transformation $\bm{\hat{W}}$
  \end{algorithmic}
\end{algorithm}

\begin{figure}[t]
	\begin{center}
		\includegraphics[width=1\columnwidth]{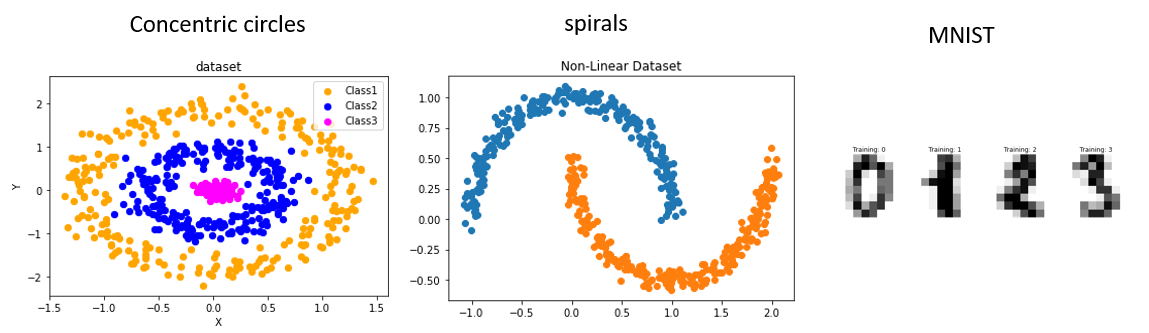} 
    \end{center}
    \caption{Classical datasets for testing performance of machine learning algorithms. The first two datasets are the classical linearly inseparable datasets. The last one is the famous MNIST digits dataset.}\label{fig:nonlineardatasets}
\end{figure}

\section{Simulation results for analog quantum variational embedding classifier}

In this section, we discuss the characterization of the performance of our algorithm for performing classification on linearly inseparable datasets. We tested our algorithm on three typical linearly inseparable datasets used for benchmarking machine learning classifiers: concentric circles, spirals, as well as MNIST images (see Fig.~\ref{fig:nonlineardatasets}). Our algorithm can do classification for all of these datasets. Here, the performance characterization is discussed in detail for concentric circles and MNIST digits, which are standard examples for machine learning benchmarking.

\subsection{Concentric circles}

The concentric circles dataset is composed of labeled points distributed in several concentric circular areas, separated by gaps~\cite{NooriBarry}. The classification task consists of training with these labeled points and using it to predict labels of unseen points. We tested our classifier on 2-circle and 3-circle cases. In our testing, we use the {\tt datasets} module from the {\tt scikit-learn} Python package to generate our dataset. We randomly sample 500/600 points as a training dataset and randomly sample 100/120 data points as a test dataset for a 2/3-circle case. We train our classifier on the training dataset, using gradient descent optimization to maximize the separation between the average density matrices corresponding to different classes. We use three $\sin$ terms in the Fourier expansion, the number of time steps is 10, and the $t_{max}$ is 2.

For a simple binary or 2-circle classification problem, the absolute loss is defined as the $L_2$ distance between the two average density matrices corresponding to the two classes. We further extend it to a multi-class case by defining a loss to maximize pair-wise distances between any two density matrices (see Algorithm~\ref{alg:architecture}). The absolute loss can be defined as the product of the pair-wise distances (see Algorithm~\ref{alg:architecture}).

After training by using a gradient descent optimizer (Adam optimizer of Pytorch, {\tt torch.optim.Adam}) to adjust the weights matrix to minimize the loss (see the 'Algorithm' section above), we test our trained classifier on the test dataset. The predicted label for each data in the test dataset is chosen to be the label of the nearest average density matrix.

\begin{figure*}[t]
	\begin{center}
		\includegraphics[width=2\columnwidth]{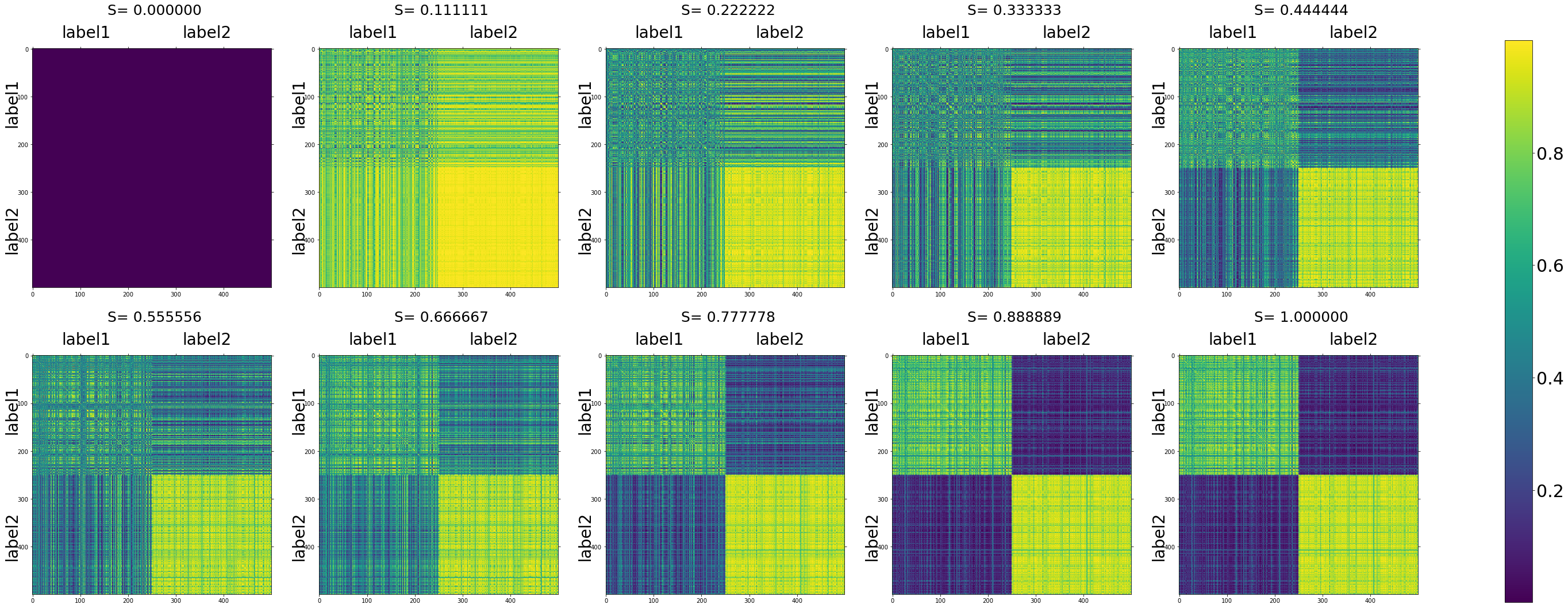} 
    \end{center}
    \caption{Time evolution of the overlap matrix of training data for binary classification on circles. The data to Hamiltonian mapping correspond to a fully trained classifier. For each time step, the evolved quantum states are calculated, and the overlap matrix is calculated. The image sequence is from $s=0$ (the top left image) to $s=1$ (the bottom right image). The brightness indicates the overlap between the embedded quantum states. Labels 1 and 2 represent the outer and inner circles, respectively.}\label{fig:Time_evolution_circles}
\end{figure*}

During training, it is observed that at the starting stages, the embedded quantum states are not clustered. Only after training do the labeled embedded quantum states start to cluster according to their labels. The overlap (squared inner product)~\cite{SethLloyd} between samples belonging to the same label is much stronger than the overlap between samples from different labels. The training results in the embedded states from the same class cluster together. The clustering of the training dataset can be visualized with the overlap matrix. Fig.~\ref{fig:Time_evolution_circles} shows the separation of different classes for the binary classification (2-circles) case. Here the h and J parameters of $H_{0}$ and $H_{1}$ are fixed, taking on values to make the initial and final Hamiltonian non-degenerate. A trained classifier is used here. The training dataset is fed into the annealer with the trained parameters. In a classification task, the final evolved quantum states of $s=1$ are used as the embedded states for the data. Here, we recorded the evolution of the quantum states at each time step. The image sequence is from $s=0$ (the top left image) to $s=1$ (the bottom right image).

\begin{figure}
     \centering
     \begin{subfigure}[t]{0.45\columnwidth}
         \caption{}
         \centering
         
         \includegraphics[width=1\columnwidth]{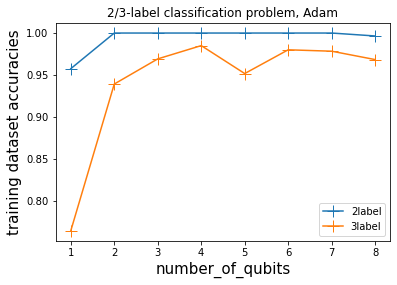}
         
         \label{fig:train_accuracy_scaling}
     \end{subfigure}
     \hfill
     \begin{subfigure}[t]{0.45\columnwidth}
         \caption{}
         \centering         
         \includegraphics[width=1\columnwidth]{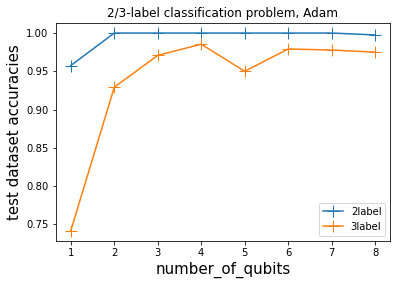}
         
         \label{fig:test_accuracy_scaling}
     \end{subfigure}

        \caption{The classification accuracy (with a trained classifier) vs. the number of qubits for multi-class classification cases (two examples -- 2-label and 3-label cases are shown here). (a) Classification accuracy vs. the number of qubits for the training dataset (dataset used for training our classifier); (b)Classification accuracy vs. the number of qubits for the test dataset (unseen dataset), directly using the classifier trained on the training dataset.}
        \label{fig:number_scaling}
\end{figure}

Next, we discuss the characterization of the performance as a function of the number of qubits. To make the dependence on the number of qubits more universal, the $h_{x_i}$, $h_{z_i}$ and $J_{ij}$ parameters of $H_{0}$ and $H_{1}$ are treated with the same footing as the Fourier coefficients here -- they are also transformed from the data by a linear transformation matrix. In Fig.~\ref{fig:number_scaling}, we show the classification accuracy vs. the number of qubits. Fig.~\ref{fig:number_scaling}(a) is the result of the training accuracy after optimizing the analog quantum computer parameters to maximize the separation of the density matrices. Fig.~\ref{fig:number_scaling}(b) is the result for the test accuracy using the same trained analog quantum computer (obtained after optimizing the analog quantum computer parameters to maximize the separation of the density matrices). For each data point, four independent random initialization (starting from four random initial guesses of the linear transformation matrix $\bm{\hat{W}}$) of the training stage are performed to obtain sufficient statistics. The mean and standard deviations of the accuracies are recorded. 

Fig.~\ref{fig:number_scaling} shows that, as the number of qubits increases,
the performance increases until it saturates and fluctuates. The saturation accuracy is $>95\%$ (see the section IV for a comparison with classical classifiers). The origin of the fluctuation may be the increased optimization
complexity as the number of qubits increases. These results show that the classification accuracies on both the training and test dataset can be generally improved by increasing the number of qubits. The similar scaling for both the training and test datasets means that our classifier has a good generalization. The performance boost as the number of qubits increases indicates that the expressivity of our classifier can be improved as the number of qubits increases. Therefore, a more complicated classification task could be handled once we add more qubits to our classifier. Whether the performance after a certain amount of iterations of training will always increase as a function of the number of qubits is a complex issue. Indeed, previous studies showed that a system with too many qubits might have barren plateaus featured with a vanishing gradient, and training or optimization of performance will become difficult in this case~\cite{nearest_centroid, barren_plateau_VQA, cerezo2021variational}. Some cures for the barren plateau have been proposed~\cite{nearest_centroid, cerezo2021variational}.

%The saturation accuracies for the training and test dataset are ($>99.5\%$, $>99.5\%$) for 2-circle case, and ($>95\%$, $>95\%$) for 3-circle case (can be further improved when trained with a different loss, see the section IV)

%The saturation accuracy is $>95\%$ (see the section IV for a comparison with classical classifiers).

\subsection{MNIST digits}
In this subsection, we tested our classifier on the MNIST digits dataset. We tested both binary classification (on digits 3 and 5) as well as multi-class classification (on digits 1, 3, and 5). Just as for the concentric circle case, the MNIST dataset is also generated by the {\tt datasets} module of the {\tt scikit-learn} package. The dimension of raw MNIST digit data from scikit-learn is an $8 \times 8$ array. The $8 \times 8$ array is reshaped into a one-dimensional array of dimension 64, which is transformed into parameters of the quantum annealer in our classifier, as described in our algorithm (see Fig.~\ref{fig:QClassifier} ). There are three $\sin$ terms in the Fourier expansion, the number of time steps is 10, and the maximum time $t_{max}$ used is 0.91 for the binary classification (on digits 3 and 5) and 0.91 for the multi-class classification (on digits 1, 3, 5). The h and J parameters are fixed here, taking on values to make the initial and final Hamiltonian non-degenerate. Only the Fourier coefficients in the schedule are transformed from the data by a linear transformation matrix. The loss used in the training stage for binary classification is $-D_{min}/r_{max}$, where $D_{min}$ is the smallest distance between centroids of different classes and $r_{max}$ is the largest spread of distance of a collection of embedded samples with respect to their corresponding centroid. 

%$std_{max}$ is the largest spread of distance to the centroid within a class
%$r_{max}$ is the largest distance of an embedded sample from its corresponding centroid. 

% $r_{max}$ is the largest spread of distance of a collection of embedded samples from their corresponding centroid

For the binary classification task on digits 3 and 5, all the digits 3 and 5 are collected, then a random $90\%$ to $10\%$ split is performed to get training and test datasets (329 samples in the collection are picked out as the training dataset and 36 samples are chosen as the test dataset). For the 3-class classification task on digits 1, 3, and 5, all the digits 1, 3, and 5 are collected, then a random $90\%$ to $10\%$ split is performed to get training and test datasets (492 samples of the collection are picked out as train dataset and 55 samples are chosen as the test dataset).

\begin{figure*}[t]
	\begin{center}
		\includegraphics[width=2\columnwidth]{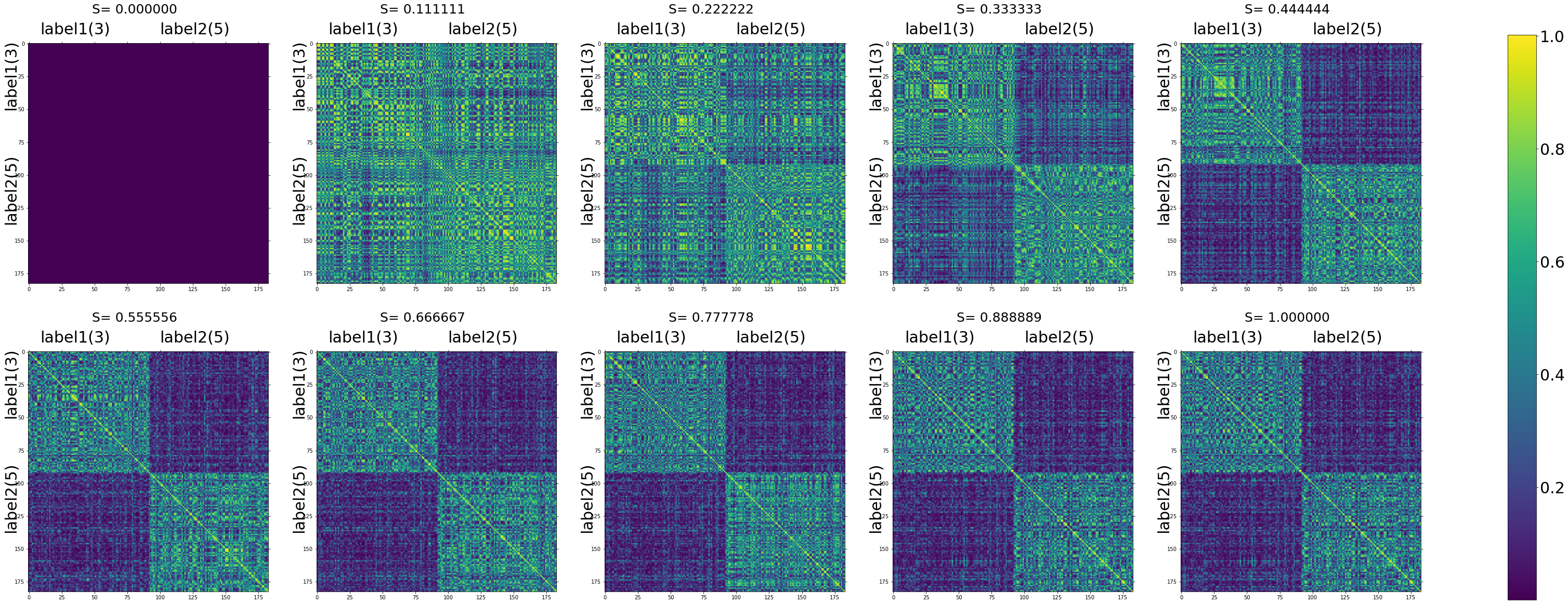} 
    \end{center}
    \caption{Time evolution of overlap matrix of training data for the binary classification on MNIST digits 3 and 5. A trained classifier is used here. The image sequence is from $s=0$ (the top left image) to $s=1$ (the bottom right image). The brightness indicates the overlap between the embedded quantum states. }\label{fig:Time_evolution_MNIST}
\end{figure*}

After training, the embedded quantum states from different labels show a clustering behavior. As usual, we visualize the clustering of embedded datasets from the same labels with the overlap matrix. Fig.~\ref{fig:Time_evolution_MNIST} shows the overlap matrix for digits classification. A clear separation is observed between different classes. A trained classifier is used here. The training dataset is fed into the annealer with the trained parameters. In a classification task, the final evolved quantum states of $s=1$ are used as the embedded states for the data. Here, we record the evolution of the quantum states at each time step. For each time step, the evolved quantum states are calculated and the overlap matrix is calculated. The image sequence is from $s=0$ (the top left image) to $s=1$ (the bottom right image). The brightness indicates the overlap between the embedded quantum states.

Next, we discuss the influence of the number of qubits on performance. For binary classification on digits 3 and 5, increasing the number of qubits, in general, can improve performance. This is shown in Table~\ref{table:MNIST35}, where the training and test accuracies for classification on MNIST digits 3 and 5 are recorded for various numbers of qubits. Each result is averaged over eight random trials, starting from eight random initial guesses of the linear transformation matrix $\bm{\hat{W}}$. For 3-class classification on digits 1, 3, and 5, increasing the number of qubits can improve the performance prominently, as shown in Table~\ref{table:MNIST135}. Here each result is also averaged over eight random trials, starting from eight random initial guesses of the linear transformation matrix $\bm{\hat{W}}$.

%the train and test accuracy is ($100\%,100\%$) for 2-circle case and ($100\%, 98.3\%$) for 3-circle case (for a simpler 2-layer neural network, the train and test accuracy is ($97.2\%, 99\%$) for a 2-circle case and ($92.83\%, 91.67\%$) for 3-circle case)

\section{Discussion}
These tests on MNIST digits and concentric circles show the power of our classifier on different datasets. In comparison with classical linear classifiers~\cite{NooriBarry}, our algorithm can significantly boost classification accuracy. The performance of our classifier is also comparable with the best classical classifiers. We tested a few best classical classifiers on the same concentric circles dataset. For a 3-layer (two 8-node hidden layers) neural-network classifier~\cite{2015NatureDL} with a Relu activation function (implemented with Pytorch~\cite{torchnnmodule}), the train and test accuracy are ($100\%,100\%$) for 2-circle case and ($99.75\%, 98.96\%$) for 3-circle case (for a simpler 2-layer neural network, the train and test accuracy are ($100\%, 100\%$) for a 2-circle case and ($99.29\%, 98.9575\%$) for 3-circle case); For a support vector machine classifier~\cite{cortes1995support, boser1992training, hofmann2008kernel, aizerman1964theoretical} (Gaussian kernel, implemented with {\tt scikit-learn}~\cite{SVMsklearn}), the train and test accuracy are ($100\%, 100\%$) for 2-circle case and ($99.167\%, 99.167\%$) for 3-circle case; For a random forest classifier~\cite{breiman2001random, ho1995random, GLoupe, geron2022hands} (with a tree depth 2, implemented with {\tt scikit-learn}~\cite{sklearnrandforest}), the train and test accuracy are ($100\%, 99\%$) for 2-circle case and ($82.3\%, 80.8\%$) for 3-circle case (for a depth-6 random forest classifier the train and test accuracy are ($100\%, 100\%$) for 2-circle case and ($99\%, 97.5\%$) for 3-circle case). As a comparison, for our classifier, the train and test accuracy are ($>99.5\%$, $>99.5\%$) for 2-circle case, and ($>95\%$, $>95\%$) for 3-circle case (can be further improved to ($>99\%$, $>99\%$) when trained with the $-D_{min}/r_{max}$ loss, $D_{min}$ is the smallest distance between centroids of different classes and $r_{max}$ is the largest spread of distance to the centroid within a class, this loss means the separation between density matrices from collections of embedded data is larger than the spread of embedded data points with respect to their centroids). 

It is worth mentioning that, the number of optimization parameters in our classifier scales linearly with $n$ (see section II). The other parameters (such as $h_{x_i}$, $h_{z_i}$ and $J_{ij}$) will not change this scaling, since they can be either fixed as constant or lead to a linear dependence on $n$ when treating them as variables. The linear scaling with $n$ makes our classifier feasible in NISQ-era. For a NISQ quantum computer with $\sim100$ qubits, the
number of control parameters ($n_{s} \times m$) is $\sim1000$. This is practical for controlling and measurement
systems in NISQ-era.

%the initial and final Ising coefficients

%The linear scaling with $n$ is a favored feature for training a large model. It makes our classifier feasible in NISQ-era. 

%For a NISQ quantum computer with $\sim100$ qubits, the
%number of control parameters is $\sim1000$, and it leads
%to a number of control parameters for flux-biasing or microwave control with
%the same magnitude. This is practical for controlling and measurement
%systems in NISQ-era.

%Regarding the other parameters: the number of Sines is constant hyper parameters thus will not contribute to the number of parameters to be optimized; The initial and final Ising coefficients ($h_{x_i}$, $h_{z_i}$ and $J_{ij}$) can be fixed as constant hyper parameters thus will not contribute to the number of parameters to be optimized in this case. If taking them as optimization parameters, the number of initial and final Ising coefficients ($h_{x_i}$, $h_{z_i}$ and $J_{ij}$) are also linear in $n$ since we use a linear chain with nearest neighbor coupling, thus they will not change the linear dependence on $n$.

In our tests, we intentionally did not optimize the hyper-parameters (such as $t_{max}$): further adjusting the hyper-parameters could further boost the performance of our classifier. Using a different loss in the training stage, such as $-D_{min}/r_{max}$ can also make the training better. 

 About the quantum part of our classifier, the quantum system used in our classifier is an implementation of a quantum annealer, but this approach could be extended to other types of quantum computers run in an analog mode. As the number of qubits increases, the nonlinearity provided by the quantum computer, in general, can not be simulated effectively with a classical computer. This could harbor a quantum advantage for quantum computation.

\begin{table}[h!]
\begin{center}
\begin{tabular}{ ||c|c|c|c|c|c|| } 
 \hline
  & 1-qubit & 2-qubit & 3-qubit & 4-qubit & 5-qubit \\ 
  \hline
 Train accuracy & 0.9206 & 0.9850 & 0.9812 & 0.9931 & 0.9942\\ 
  \hline
 Test accuracy & 0.8590 & 0.9359 & 0.9519 & 0.9679 & 0.9744\\ 
 \hline
\end{tabular}
\end{center}
\caption{Performance results for binary classification on MNIST digits 3 and 5. The train accuracy and test accuracy (average results from 8 random trials) for quantum annealers with various numbers of qubits are recorded.}
\label{table:MNIST35}
\end{table}

\begin{table}[h!]
\begin{center}
\begin{tabular}{ ||c|c|c|c|| } 
 \hline
   & 2-qubit & 4-qubit & 5-qubit \\ 
  \hline
 Training accuracy & 0.9098 & 1 & 1\\ 
  \hline
 Test accuracy & 0.7864 &  0.9454 & 0.9523\\ 
 \hline
\end{tabular}
\end{center}
\caption{Performance results for a 3-class classification on the MNIST digits 1, 3, and 5. The train accuracy and test accuracy (average results from 8 random trials) for quantum annealers with various numbers of qubits are recorded. We can see that increasing the number of qubits can increase the train accuracy or test accuracy prominently.}
\label{table:MNIST135}
\end{table}

\section{Summary}

We proposed an analog quantum variational embedding classifier with a focus on an implementation based on a quantum annealer. The nonlinear mapping of the classical data to a high dimensional density matrix is realized with an analog quantum computer. The classical data is transformed into the parameters of the analog quantum computer by a linear transformation, which implies that the nonlinearity needed for a nonlinear classification problem arises purely from the analog quantum computer due to the nonlinear dependence between the final quantum state and the control parameters of the Hamiltonian. By using a metric based on density matrix from a collection of training dataset~\cite{SethLloyd}, our algorithm can handle a general classification problem. Moreover, our classifier handles both binary and multi-class classification tasks. We demonstrate the effectiveness of our algorithm for performing binary and multi-class classification on linearly inseparable datasets. Our algorithm performs much better than a classical linear classifier. The performance of our classifier is also comparable with that of the best classical classifiers. The dependence of performance on the number of qubits shows that increasing the number of qubits can improve performance until the performance saturates and fluctuates.  In addition, the number of optimization parameters of our classifier scales linearly with the number
of qubits. This linear scaling is an advantage when comparing with classical neural network whose number of training parameters
scales quadratically ($O(n^2)$) with the number of nodes. Our algorithm presents the possibility of using current and near-term quantum annealers for solving practical machine-learning problems, and it could also be useful to explore quantum advantage in quantum machine learning. 

As a prospect, in the future, topics such as the performance of our classifier with other types of Hamiltonians, the expressivity of the analog quantum computer, the universality of the nonlinearity, the experimental realization on actual quantum computers, etc., can be investigated.

\begin{acknowledgments}

This material is based upon work supported by the Defense Advanced Research Projects Agency (DARPA) under agreement No.HR00112109969

\end{acknowledgments}

\newpage

%\bibliography{QClassifier}

\providecommand{\noopsort}[1]{}\providecommand{\singleletter}[1]{#1}%

\end{document}